\documentclass[epsfig,graphics,twocolumn,floatfix,mathbbm,prl,nobibnotes]{revtex4}
\usepackage{graphicx}
\usepackage{hyperref} 
\usepackage{epstopdf}
\usepackage{amsmath}

\begin{document}

\newcommand{\Nd}{Nd$^{3+}$}
\newcommand{\Y}{Y$^{3+}$}
\newcommand{\NdYVO}{Nd$^{3+}$:YVO$_{4}$}
\newcommand{\gstate}{$^{4}$I$_{9/2}$}
\newcommand{\estate}{$^{4}$F$_{3/2}$}
\newcommand{\trans}{$^{4}$I$_{9/2}$ $\rightarrow$ $^{4}$F$_{3/2}$}

\title[a]{High precision measurement of the Dzyaloshinsky-Moriya interaction between two rare-earth ions in a solid}

\author{Cyril Laplane$^{1}$}
\thanks{These authors contributed equally to this work.}
\author{Emmanuel Zambrini Cruzeiro$^{1}$}
\thanks{These authors contributed equally to this work.}
\author{Florian Fr\"{o}wis$^{1}$}
\author{Philippe Goldner$^{2}$}
\author{Mikael Afzelius$^{1}$}
\email{mikael.afzelius@unige.ch}

\address{$^{1}$Group of Applied Physics, University of Geneva, CH-1211 Geneva 4, Switzerland}
\address{$^{2}$PSL Research University, Chimie ParisTech, CNRS, Institut de Recherche de Chimie Paris, 75005, Paris, France}

\date{\today}

\begin{abstract}
We report on a direct measurement of the pair-wise anti-symmetric exchange interaction, known as the Dzyaloshinsky-Moriya interaction (DMI), in a Nd$^{3+}$-doped YVO$_4$ crystal. To this end we introduce a broadband electron spin resonance technique coupled with an optical detection scheme which selectively detects only one Nd$^{3+}$-Nd$^{3+}$ pair. Using this technique we can fully determine the spin-spin coupling tensor, allowing us to experimentally determine both the strength and direction of the DMI vector. We believe that this ability to fully determine the interaction Hamiltonian is of interest for studying the numerous magnetic phenomena where the DMI interaction is of fundamental importance, including multiferroics. We also detect a singlet-triplet transition within the pair, with a highly suppressed magnetic-field dependence, which suggests that such systems could form singlet-triplet qubits with long coherence times for quantum information applications.
\end{abstract}

\maketitle

The detailed understanding of strongly interacting rare-earth (RE) ions is important for describing magnetic phenomena in many solid-state materials. The anti-symmetric exchange interaction, known as Dzyaloshinsky-Moriya interaction (DMI) \cite{Dzyaloshinsky1958,Moriya1960}, is particularly important for a wide range of magnetic phenomena. The DMI causes twisting of spins (canting), which can lead to weak ferromagnetism in antiferromagnetic materials \cite{Moriya1960} or the formation of magnetic skyrmions \cite{Muhlbauer2009}. DMIs also play fundamental roles for magnetoelectric effects in multiferroics \cite{Katsura2005,Sergienko2006} and for molecular magnetism \cite{Zhao2003}. It has proven difficult, however, to experimentally measure both the magnitude and direction of the usually weak DMI, although some recent progress have been made in special cases. Both the DMI direction and strength were measured in the weak ferromagnet FeBO$_3$ using a novel technique based on X-ray diffraction interference \cite{Dmitrienko2014}. The DMI vector could also be determined in a thin magnetic film using Brillouin light scattering \cite{Nembach2015}. The \textit{uniform} DMI in spin chains was recently measured using electron spin resonance (ESR) techniques \cite{Gangadharaiah2008,Povarov2011,Fayzullin2013,Haelg2014}. The magnitude of the DMI has also been extracted from neutron inelastic scattering \cite{Mena2014}.

In this work we fully characterize the general spin-spin interaction Hamiltonian $\textbf{H}_{int} = \textbf{S}_1 \cdot \tilde{J} \cdot \textbf{S}_2$ of two nearest-neighbour rare-earth ion spins, with $S_{1,2}=1/2$. We make no \textit{a priori} assumptions of the nature, nor the symmetry of the $\tilde{J}$ coupling tensor. To measure $\tilde{J}$ we have developed a broadband ESR technique, coupled with optical detection (OD-ESR), which allows us to measure the energy spectrum of the coupled ions from zero magnetic field all through to the high-field limit. Consequently the Zeeman energy ranges from being a small perturbation of the $\tilde{J}$ coupling, to being the dominant energy scale. In conventional ESR spectroscopy only the latter region can be probed, which prevents one to determine all elements of $\tilde{J}$ \cite{Blanchard2015}. With our technique we can fully determine $\tilde{J}$, where the anti-symmetric part of the tensor is the DMI. The OD is crucial, because it allows us to selectively detect the ESR signal of only a certain pair of the closest rare-earth ion neighbours. The narrow spin resonances (around 10 MHz) also makes our method sensitive to interaction energies below 0.1 $\mu$eV, allowing high precision measurements of the weak DMI.

The sample is a uniaxial \NdYVO\ crystal, where \Nd\ ions substitute some of the \Y\ ions (0.1\% \Nd\ doping) in a crystallographic site of D$_{2d}$ point symmetry. Due to the site symmetry the \Nd\ electronic ground state $^{4}$I$_{9/2}$ splits into five Kramers doublets, each described as effective $S=1/2$ spins. At our working temperature of around 3 K all ions are in the Kramers doublet with lowest energy. One can therefore model the pair-wise interaction between these ions as an interaction between two identical spin $S=1/2$ particles.

\begin{figure}
	\centering
	\includegraphics[width=0.5\textwidth]{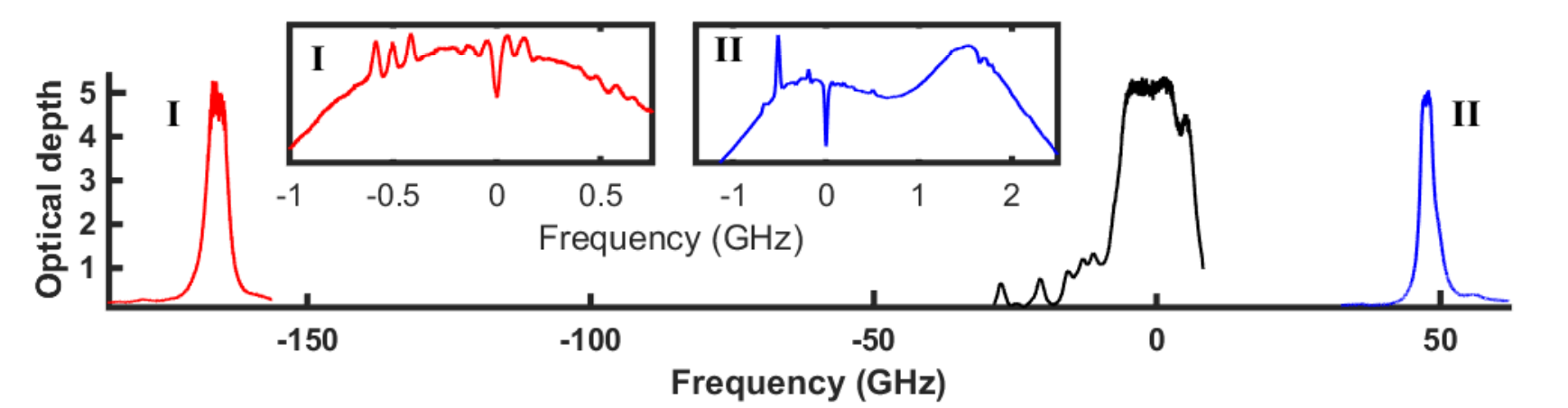}
        \caption{Optical absorption spectra showing one red (I) and one blue (II) detuned satellite lines, with energy shifts of -166 GHz (I) and +48 GHz (II) with respect to the central line due to isolated \Nd\ ions. Note that the measured optical depth saturates for the central line. The inset shows the spectral hole burning spectra (at zero magnetic field). The hole is at zero relative frequency. The anti-holes are due to distinct ground-state structures associated with each line.}
 	\label{fig:absorptionspectra}
\end{figure}

To optically detect the \Nd\ ions in the ground state we measure the absorption on the \trans\ transition. The crystal is cooled down to about 3 K in order to obtain narrow optical absorption lines \cite{Hastings-Simon2008}. Isolated \Nd\ ions (not forming pairs) absorb light at 879.66 nm (the central line). At our doping concentration one can also observe satellite lines, both red and blue detuned from the central line. It has already been suggested that these lines originate from coupled \Nd\ pairs \cite{Guillot-Noel2000}. In Figure \ref{fig:absorptionspectra} we show the absorption spectra of the two strongest of these lines, one red detuned and one blue detuned. These lines are surprisingly narrow, with linewidths (around 2 GHz) comparable to weakly doped (10 ppm) \NdYVO\ crystals \cite{Hastings-Simon2008}. The narrow linewidths allow us to optically probe these lines independently without spectral interference.

In Ref. \cite{Guillot-Noel2000} it was suggested that pairs consisting of one red detuned and one blue detuned satellite line were due to strong isotropic exchange coupling. According to that model the exchange coupling magnitude would simply be the spectral separation between the lines, which ranges from 24 (0.1 meV) to 147 GHz (0.6 meV) for the satellite lines studied in Ref. \cite{Guillot-Noel2000}. That model also predicts that all the red-detuned absorption lines originate from the singlet $S=0$ eigenstate of the isotropic exchange Hamiltonian, i.e. the red detuned line should be a single absorption line without fine structure.

To investigate a possible fine structure we employ spectral hole burning (SHB) spectroscopy. This is a well-established optical technique to probe closely spaced transitions that are otherwise hidden within the inhomogeneous broadening \cite{Macfarlane1987a}. By burning a hole for much longer than the radiative lifetime of the excited state one can increase the population, with respect to thermal equilibrium, in other ground-state spin levels. This results in increased absorption, or anti-holes, at frequencies related to the ground-state splittings. A condition for observing anti-holes is a slow ground-state spin relaxation rate \cite{Hastings-Simon2008a,Afzelius2010b}, which often can be achieved in RE ion doped crystals at low temperatures.

In Figure \ref{fig:absorptionspectra} we show the SHB spectra for both satellite lines. The anti-holes in the red-detuned line clearly show that it cannot be a singlet ground state. Moreover, the two distinctly different SHB spectra indicate that some type of interactions cause both a global shift of the optical line and a complicated ground-state spin manifold. This is supported by recent OD-NMR measurements in a Ce$^{3+}$:EuCl$_3\cdot$6H$_2$O crystal  \cite{Ahlefeldt2013a,Ahlefeldt2013}, where it was shown that the Ce$^{3+}$-Eu$^{3+}$ pair-formation causes a large optical shift due to electronic interactions and a site-specific NMR spectrum due to magnetic dipole-dipole interactions between the ions. In the following we fully characterize the ESR spectrum of the red-detuned line and we show that it is due to pairing of \Nd\ ions through a weak anti-symmetric exchange interaction.

Conventional ESR spectroscopy is narrowband (at a fixed frequency, e.g. the X-band) due to the microwave cavity that is used to increase the detection sensitivity. Using OD-ESR, however, the high sensitivity of the optical detection makes broadband ESR possible, in principle. It remains a problem, however, to efficiently excite a sample over a large bandwidth. Recently ensembles of spins have been coupled to superconducting co-planar resonators (CPR) in the microwave regime (5-10 GHz typically) \cite{Kubo2010,Schuster2010,Bushev2011}. Also, spin resonances due to Cr$^{3+}$ in ruby have been detected by measuring microwave absorption using a co-planar transmission line (CPTL) \cite{Schuster2010,Clauss2013,Wiemann2015}, demonstrating the usefulness of these novel techniques for broadband ESR spectroscopy.

Here we combine a CPTL for broadband ESR excitation and optical detection for high sensitivity and spectral selectivity. The YVO$_4$ crystal was placed on top of a 10 mm long copper CPTL on an AlO$_2$ substrate, with a $d$=900 $\mu$m distance between the ground plate and the transmission line. The penetration depth (in the sample) of the magnetic field due to the CPTL is expected to be roughly $d$. The  laser beam is aligned along the CPTL and close enough to allow optical and ESR excitation of a common volume. A superconducting magnet can apply a static magnetic field $B$ perpendicular to the CPTL. An ESR resonance is detected by monitoring the optical spectral hole created by the SHB (cf. inset in Figure \ref{fig:absorptionspectra}) when applying a radio-frequency (RF) pulse at frequency $\omega_{RF}$ through the CPTL. If the RF pulse is in resonance with a spin transition within the ground-state it will rethermalize some of the spins, thereby reducing the depth of the spectral hole.

\begin{figure*}
	\centering
	\includegraphics[width=1\textwidth]{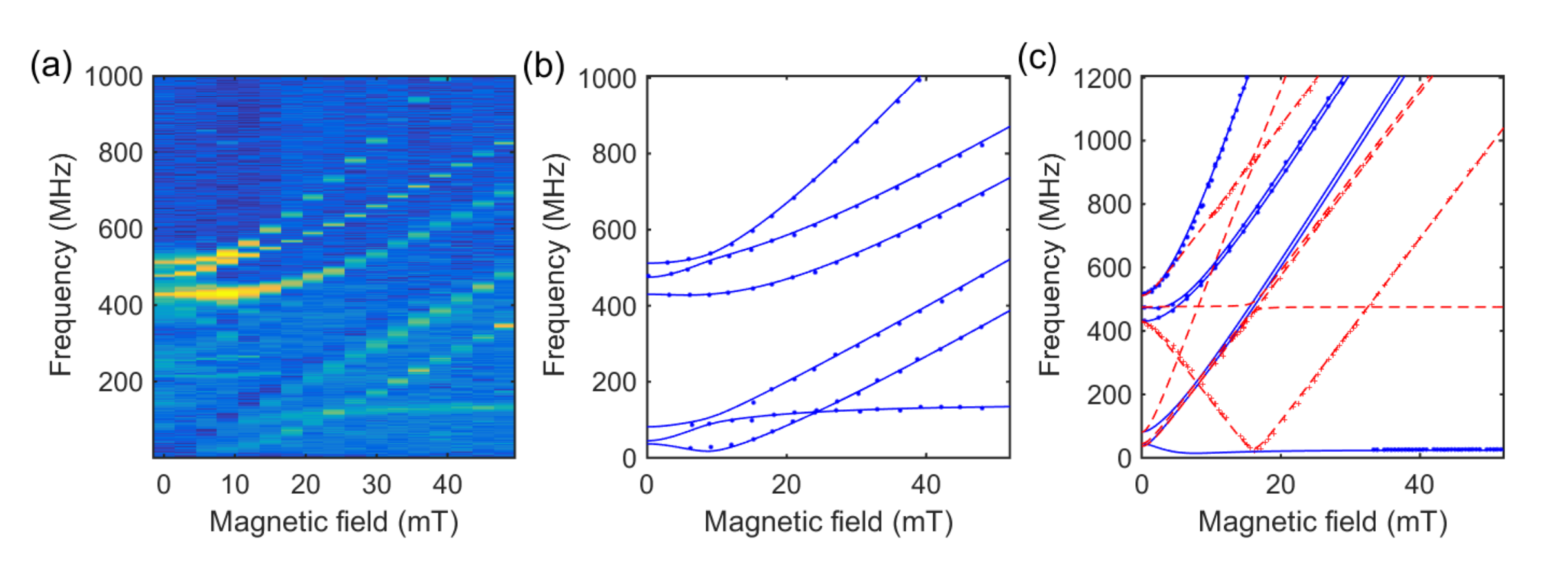}
	\caption{(a) OD-ESR measurements with the field along the $c$ axis. (b) Spin resonances (blue dots) extracted from (a), and the theoretical spectrum (blue lines) calculated from the fitted spin Hamiltonian $\textbf{H}$ (see text for details). (c) Spin resonances extracted from OD-ESR measurements with the field along the $a$ axis. The detected resonances are divided into class 1 (blue dots) and class 2 (red squares) according to the $g$ factor (see text for details). The corresponding theory spectra are shown as solid-blue and dashed-red lines.}
	\label{fig:ODESRspectra}
\end{figure*}

In Figure \ref{fig:ODESRspectra}a we show the OD-ESR spectrum for a $B$ field applied along the crystal $c$-axis. The resonance frequencies extracted from the 2D images are shown in Figure \ref{fig:ODESRspectra}b. At zero field we clearly observe three strong resonances at 430, 475 and 512 MHz. For low fields these resonances show a non-linear field dependence, while at higher fields the resonances tend to a more linear dependence. Starting from around 10-15 mT, three weaker resonances appear at 100 MHz or below. At high fields, four out of the six resonance lines have a linear dependence corresponding to a $g$ factor of $g_c=0.768$, compared to 0.915 for non-interacting \Nd\ ions \cite{Guillot-Noel1999}. A strong evidence of coupled ions is the single line that has exactly twice the linear dependence $2g_c$ at high fields. This resonance couples the $m_S = \pm 1$ levels within an effective $S=1$ state. Similarly we observe one line which tends to a very weak, linear $B$ dependence of only 112 $\pm$ 4 MHz/T. This resonance couples the two $m_S = 0$ levels in the effective $S=0$ and $S=1$ states, respectively. Altogether the six resonances lines, which can be explained by only four levels, and their field dependencies strongly suggest two coupled spin 1/2 ions. To our knowledge all these features have not previously been observed in a single ESR experiment.

\begin{figure}
	\centering
	\includegraphics[width=0.5\textwidth]{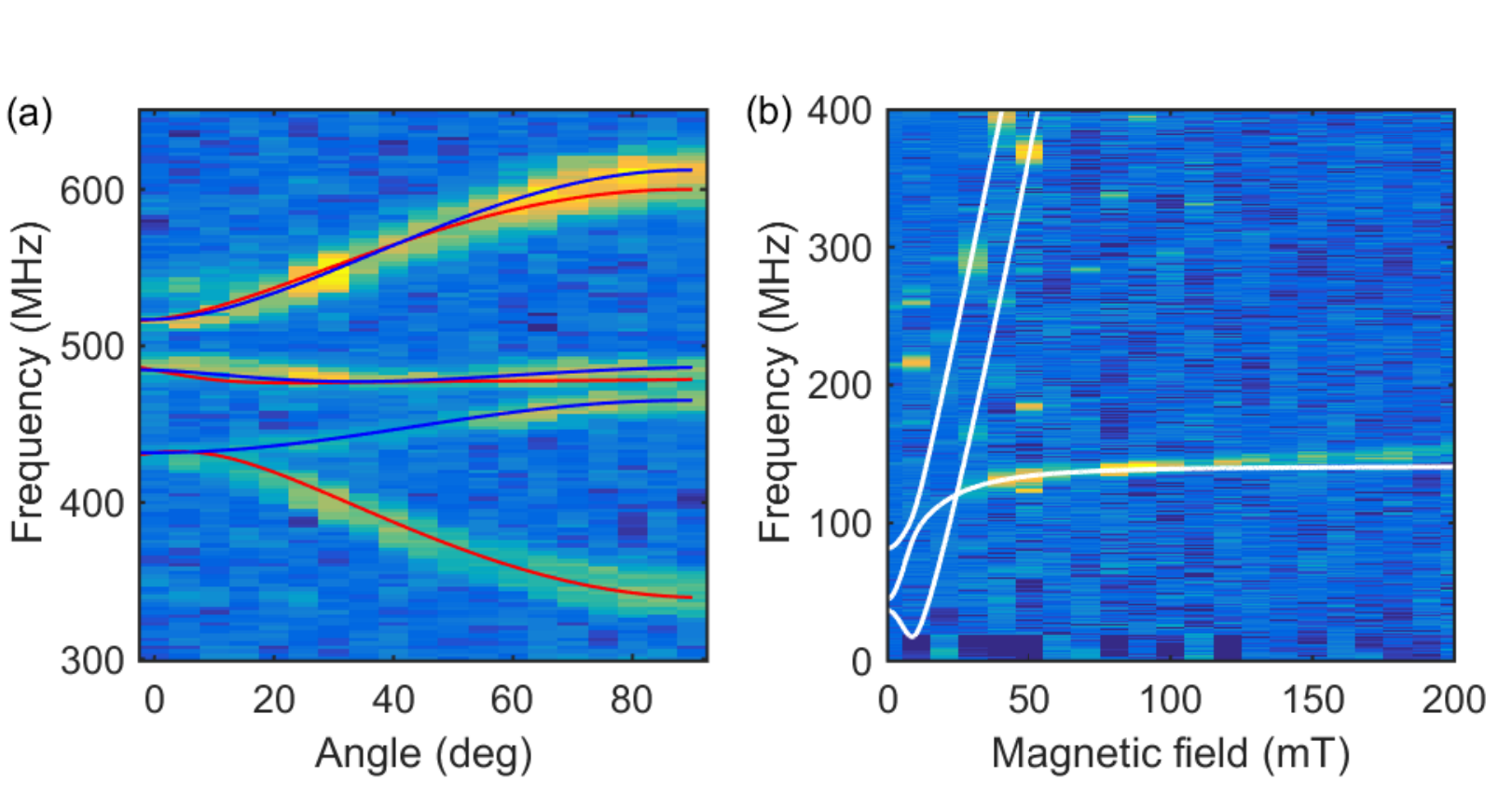}
	\caption{(a) OD-ESR measurements as a function of angle to the $c$ axis, in the $a$-$c$ plane, for a fixed field strength of 4.3 mT. The model prediction is superimposed, with class 1 in blue and class 2 in red. (b) OD-ESR measurements with the field along the $c$ axis. The weakly field dependent singlet-triplet transition is detected up to a field of 200 mT. The model prediction is shown as white lines.}
	\label{fig:ODESRspectraangle}
\end{figure}

Since we expect the interaction to be highly anisotropic we also performed a measurement with $B$ along the $a$-axis. The resonance frequencies extracted from the 2D images (see Supplemental Material) are shown in Figure \ref{fig:ODESRspectra}c. The spectrum is distinctly different as compared to the c-axis, a result of the anisotropy. A close inspection shows that the two zero-field resonances at 430 and 512 MHz split into two resonance lines (see also Fig.\ref{fig:ODESRspectraangle}a where the split of the 430 MHz line is clearly visible for any field not parallel with the $c$ axis). To understand this behaviour we need to consider the possible orientations of a given pair of ions with respect to the crystallographic $a$,$b$,$c$ axes. Indeed, for a given crystallographic pair of nearest-neighbour ions there exist several possible orientations, such that these pairs become in-equivalent due to the applied magnetic field. In our case the split into two sets of resonance lines for any field not along the $c$-axis implies that our pair can be aligned either along the $a$ or $b$ axis (see Supplemental Material). In the following we refer to these two groups of lines as classes 1 and 2. We emphasize that these resonance lines stem from identical pairs described by the same interaction $\tilde{J}$, only their orientation with respect to the external field induce the split.

In the high-field limit the two classes have slightly different $g$ factors in the $a$-$b$ plane, $g_a=2.067$ and $g_b=2.516$. For non-interacting \Nd\ ions the $g$ factor is axially symmetric (i.e. $g_a=g_b$).  This breaking of the axial symmetry is probably induced by the pair formation itself, which would affect the $g$ factor through a symmetry-breaking of the crystal field Hamiltonian. We also observed a field-insensitive  transition for class 1, at around 25 MHz at high fields, with a linear $B$ dependence of 46$\pm$6 MHz/T and a linewidth of only 4 MHz.

The interaction Hamiltonian $\textbf{H}_{int}$ can now be fitted in order to obtain the best agreement with all the spin resonance lines observed in Figure \ref{fig:ODESRspectra} . We then need to consider the total Hamiltonian including the Zeeman interaction of each ion $\textbf{H}=\textbf{H}_{int} + \textbf{B} \cdot \tilde{g} \cdot \textbf{S}_1 + \textbf{B} \cdot \tilde{g} \cdot \textbf{S}_2$, where $\textbf{B}$ is the magnetic field vector and $\tilde{g}$ a matrix. We here assume that $\tilde{g}$ is diagonal in the reference frame $a,b,c$, with the principal values 2.067, 2.516, 0.768 extracted from the high-field data as discussed above. To calculate class 2 resonance lines the $\tilde{g}$ and $\tilde{J}$ matrices are both rotated by 90 degrees around the $c$ axis, while keeping the $B$ field direction fixed.

We found that the complete set of resonances could not be fitted assuming a symmetric interaction matrix $\tilde{J}$. Only by allowing an anti-symmetric DMI interaction could an excellent agreement be obtained, as seen in Figures \ref{fig:ODESRspectra}b-c. Many fine details of the field dependencies are reproduced, such as the avoided-level crossing between the 475 MHz and 512 MHz lines at around 10 mT. Also the curvature of the resonance line of class 2 that has a turning point at around 25 MHz for a field of 17 mT. To test our model we also performed an angle scan, with a fixed magnetic field strength. As shown in Figure \ref{fig:ODESRspectraangle}a the predicted resonance lines show excellent agreement with the experimental spectrum, without any additional tuning of parameters. The non-linear behaviour in this field-region represents a severe test for the model.

We now turn to the analysis of the DMI interaction. The fitted coupling tensor is $\tilde{J}$=[$451$ $-132$ $-31$; $-40$ $-271$ $345$; $-19$ $-325$ $-402$] (in MHz). Since $\tilde{J}$ describes one of the two possible orientations of this pair in the $a$-$b$ plane (class 1), the following analysis applies to this orientation. Now, $\tilde{J}$ can be split into a symmetric $\tilde{J}_S$ and anti-symmetric part $\tilde{J}_{AS}$, such that $\tilde{J}=\tilde{J}_{S}+\tilde{J}_{AS}$. The anti-symmetric part can further be expressed as $\tilde{J}_{AS} = \textbf{D} \cdot (\textbf{S}_1 \times \textbf{S}_2)$, where $\textbf{D}$ is the DMI vector. As shown by Moriya \cite{Moriya1960} the direction of $\textbf{D}$ is constrained by symmetry considerations of the interacting pair, such that only two of the five nearest-neighbour pairs could possess a DMI. Furthermore, only one of these is also consistent with the resonance split discussed above. This pair is aligned along either the $a$ or $b$ axes, with a distance of 7.12 \AA. According to Moriya the corresponding DMI vectors should point along either $b$ or $a$, perpendicularly to the inter-pair vector.

\begin{figure}[b]
	\centering
	\includegraphics[width=0.45\textwidth]{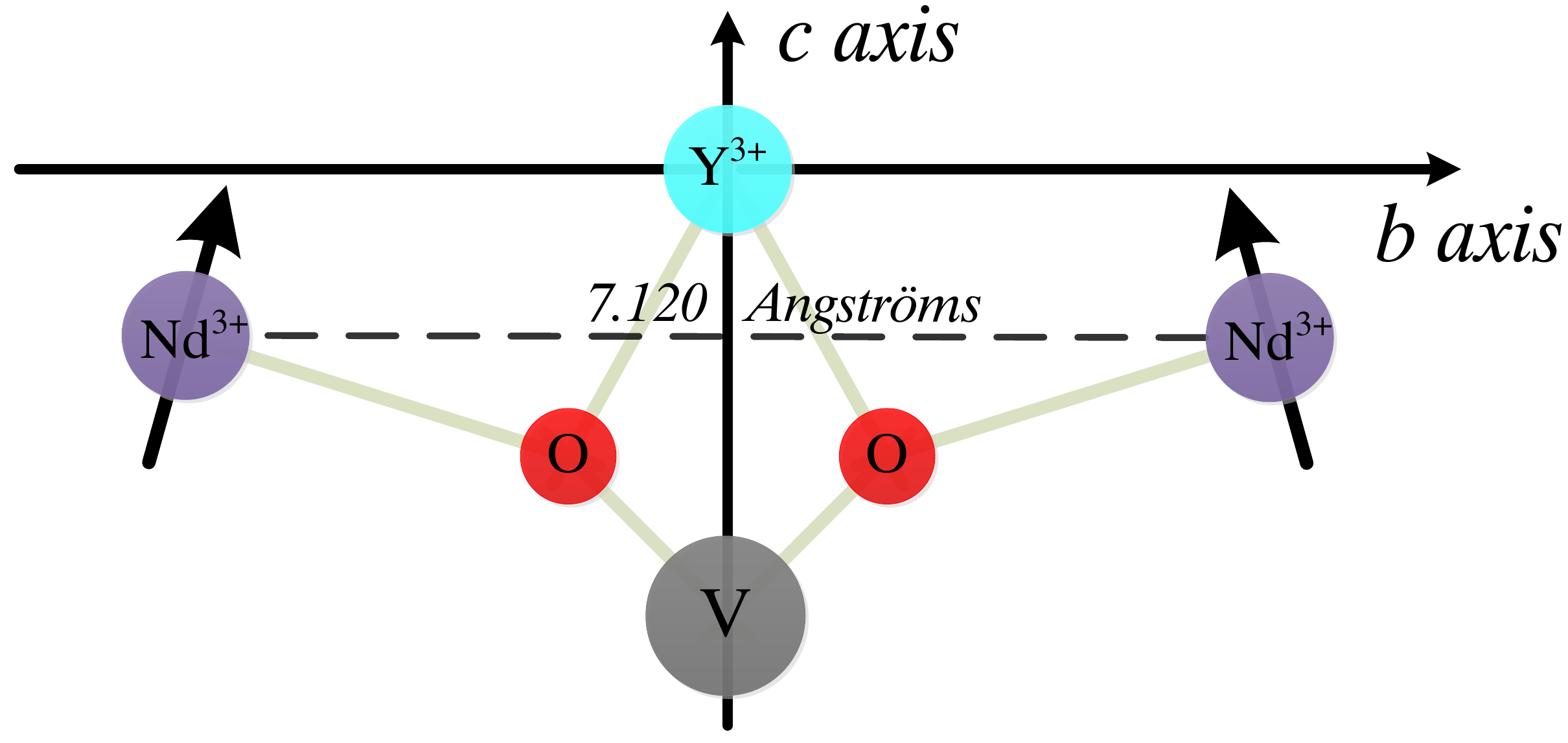}
        \caption{Shown is a schematic of the relative positions of the interacting \Nd\ ions separated by 7.12 \.{A}. The oxygen ions, offset from the line joining the \Nd\ ions, are likely to mediate the anti-symmetric exchange interaction.}
 	\label{fig:pair}
\end{figure}

The DMI vector calculated using $\tilde{J}_{AS}$ is $\textbf{D} = [335, 6, -47]$ MHz, which points along $a$ within 10$^{\circ}$, as expected for the symmetry. Consequently, the second possible pair orientation has the same DMI vector, but aligned close to $b$ (see discussion above). The consistency with Moriya's symmetry arguments further strengthens the confidence in the determined coupling tensor $\tilde{J}$. For this particular pair, the presence of two oxygen atoms in between the \Nd ions, slightly offset with respect to the inter-pair vector, suggests an anti-symmetric exchange interaction mediated by superexchange through the oxygen atoms, see Figure \ref{fig:pair}. The magnitude of the DMI is weak, about 1.4 $\mu$eV, likely due to the large separation. Yet our measurements allow us to determine its magnitude and direction with precision. We emphasize that the anti-symmetric interaction energy is comparable in magnitude to the symmetric part of the total spin-spin interaction, which is a sum of symmetric exchange and dipole-dipole interactions (see Supplemental Material).

The magnetic properties of NdVO$_4$ have not been well-studied. We note, however, that magnetic susceptibility measurements along the \textit{a}-axis have indicated a weak ferromagnetism that was thought to be due to spin canting in NdVO$_4$ \cite{Suzuki1980,Suzuki1983}. Our observation of a DMI interaction along the $a$/$b$ axes could be the underlying microscopic interaction behind the weak ferromagnetism. This indicates an interesting avenue for future measurements and comparisons with theoretical modelling.

Our work also points to an interesting possibility of directly measuring a possible magnetoelectric effect on the microscopic DMI. Using our experimental technique the size and direction of the DMI vector could be measured while applying an electric field. Such experiments might might shed light on the role of DMI in multiferroics \cite{Katsura2005,Sergienko2006}.

Interacting RE ions in crystals are not only interesting for magnetic phenomena in solids, but also for their role in quantum information applications \cite{Ahlefeldt2013,Farr2015}. RE ions are currently investigated as ensembles of ions for quantum memories for light\cite{Bussieres2013,RiedmattenAfzeliusChapter2015} and as single-ion qubits \cite{Rippe2008,Kolesov2012,Yin2013,Siyushev2014,Utikal2014,Xia2015,Eichhammer2015}. We argue that exchange-coupled RE ions can be used as singlet-triplet qubits with strongly reduced sensitivity to the dephasing spin bath, analogous to singlet-triplet qubits in two-electron quantum dots \cite{Petta2005}. Experimentally we have observed two singlet-triplet transitions (Figs \ref{fig:ODESRspectra}b-c, \ref{fig:ODESRspectraangle}b) having weak linearities comparable to nuclear spin states. Using a perturbation calculation one can show that the singlet-triplet transition is increasingly decoupled from the dephasing spin bath with an increasing bias $B$ field (see Supplemental Material). This could provide an alternative to zero first-order Zeeman (ZEFOZ) transitions \cite{McAuslan2012}, which require extremely precise alignment of the $B$ field. Future studies using pulsed OD-ESR measurements should be made to determine the spin coherence time for these transitions.

In conclusion we have demonstrated a novel OD-ESR approach that allows us to make high-precision measurements of the spin-spin coupling over a wide range of energies. We have shown that it allows us to directly measure both the magnitude and direction of the anti-symmetric DMI exchange interaction, which is of great interest for understanding numerous magnetic phenomena. We also suggest that exchange-coupled ions could provide singlet-triplet qubits that are naturally decoupled from the dephasing spin bath, which are interesting for quantum information applications, such as quantum memories for photons.

We acknowledge fruitful discussions with F\'{e}lix Bussi\`{e}res and Pavel Sekatski. 




\begin{thebibliography}{10}


\end{thebibliography}

\begin{thebibliography}{41}%
\makeatletter
\providecommand \@ifxundefined [1]{%
 \@ifx{#1\undefined}
}%
\providecommand \@ifnum [1]{%
 \ifnum #1\expandafter \@firstoftwo
 \else \expandafter \@secondoftwo
 \fi
}%
\providecommand \@ifx [1]{%
 \ifx #1\expandafter \@firstoftwo
 \else \expandafter \@secondoftwo
 \fi
}%
\providecommand \natexlab [1]{#1}%
\providecommand \enquote  [1]{``#1''}%
\providecommand \bibnamefont  [1]{#1}%
\providecommand \bibfnamefont [1]{#1}%
\providecommand \citenamefont [1]{#1}%
\providecommand \href@noop [0]{\@secondoftwo}%
\providecommand \href [0]{\begingroup \@sanitize@url \@href}%
\providecommand \@href[1]{\@@startlink{#1}\@@href}%
\providecommand \@@href[1]{\endgroup#1\@@endlink}%
\providecommand \@sanitize@url [0]{\catcode `\\12\catcode `\$12\catcode
  `\&12\catcode `\#12\catcode `\^12\catcode `\_12\catcode `\%12\relax}%
\providecommand \@@startlink[1]{}%
\providecommand \@@endlink[0]{}%
\providecommand \url  [0]{\begingroup\@sanitize@url \@url }%
\providecommand \@url [1]{\endgroup\@href {#1}{\urlprefix }}%
\providecommand \urlprefix  [0]{URL }%
\providecommand \Eprint [0]{\href }%
\providecommand \doibase [0]{http://dx.doi.org/}%
\providecommand \selectlanguage [0]{\@gobble}%
\providecommand \bibinfo  [0]{\@secondoftwo}%
\providecommand \bibfield  [0]{\@secondoftwo}%
\providecommand \translation [1]{[#1]}%
\providecommand \BibitemOpen [0]{}%
\providecommand \bibitemStop [0]{}%
\providecommand \bibitemNoStop [0]{.\EOS\space}%
\providecommand \EOS [0]{\spacefactor3000\relax}%
\providecommand \BibitemShut  [1]{\csname bibitem#1\endcsname}%
\let\auto@bib@innerbib\@empty
\bibitem [{\citenamefont {Dzyaloshinsky}(1958)}]{Dzyaloshinsky1958}%
  \BibitemOpen
  \bibfield  {author} {\bibinfo {author} {\bibfnamefont {I.}~\bibnamefont
  {Dzyaloshinsky}},\ }\href
  {http://www.sciencedirect.com/science/article/pii/0022369758900763}
  {\bibfield  {journal} {\bibinfo  {journal} {Journal of Physics and Chemistry
  of Solids}\ }\textbf {\bibinfo {volume} {4}},\ \bibinfo {pages} {241}
  (\bibinfo {year} {1958})}\BibitemShut {NoStop}%
\bibitem [{\citenamefont {Moriya}(1960)}]{Moriya1960}%
  \BibitemOpen
  \bibfield  {author} {\bibinfo {author} {\bibfnamefont {T.}~\bibnamefont
  {Moriya}},\ }\href {\doibase 10.1103/PhysRev.120.91} {\bibfield  {journal}
  {\bibinfo  {journal} {Phys. Rev.}\ }\textbf {\bibinfo {volume} {120}},\
  \bibinfo {pages} {91} (\bibinfo {year} {1960})}\BibitemShut {NoStop}%
\bibitem [{\citenamefont {M\"{u}hlbauer}\ \emph {et~al.}(2009)\citenamefont
  {M\"{u}hlbauer}, \citenamefont {Binz}, \citenamefont {Jonietz}, \citenamefont
  {Pfleiderer}, \citenamefont {Rosch}, \citenamefont {Neubauer}, \citenamefont
  {Georgii},\ and\ \citenamefont {B\"{o}ni}}]{Muhlbauer2009}%
  \BibitemOpen
  \bibfield  {author} {\bibinfo {author} {\bibfnamefont {S.}~\bibnamefont
  {M\"{u}hlbauer}}, \bibinfo {author} {\bibfnamefont {B.}~\bibnamefont {Binz}},
  \bibinfo {author} {\bibfnamefont {F.}~\bibnamefont {Jonietz}}, \bibinfo
  {author} {\bibfnamefont {C.}~\bibnamefont {Pfleiderer}}, \bibinfo {author}
  {\bibfnamefont {A.}~\bibnamefont {Rosch}}, \bibinfo {author} {\bibfnamefont
  {A.}~\bibnamefont {Neubauer}}, \bibinfo {author} {\bibfnamefont
  {R.}~\bibnamefont {Georgii}}, \ and\ \bibinfo {author} {\bibfnamefont
  {P.}~\bibnamefont {B\"{o}ni}},\ }\href@noop {} {\bibfield  {journal}
  {\bibinfo  {journal} {Science}\ }\textbf {\bibinfo {volume} {323}},\ \bibinfo
  {pages} {915} (\bibinfo {year} {2009})}\BibitemShut {NoStop}%
\bibitem [{\citenamefont {Katsura}\ \emph {et~al.}(2005)\citenamefont
  {Katsura}, \citenamefont {Nagaosa},\ and\ \citenamefont
  {Balatsky}}]{Katsura2005}%
  \BibitemOpen
  \bibfield  {author} {\bibinfo {author} {\bibfnamefont {H.}~\bibnamefont
  {Katsura}}, \bibinfo {author} {\bibfnamefont {N.}~\bibnamefont {Nagaosa}}, \
  and\ \bibinfo {author} {\bibfnamefont {A.~V.}\ \bibnamefont {Balatsky}},\
  }\href {\doibase 10.1103/PhysRevLett.95.057205} {\bibfield  {journal}
  {\bibinfo  {journal} {Phys. Rev. Lett.}\ }\textbf {\bibinfo {volume} {95}},\
  \bibinfo {pages} {057205} (\bibinfo {year} {2005})}\BibitemShut {NoStop}%
\bibitem [{\citenamefont {Sergienko}\ and\ \citenamefont
  {Dagotto}(2006)}]{Sergienko2006}%
  \BibitemOpen
  \bibfield  {author} {\bibinfo {author} {\bibfnamefont {I.~A.}\ \bibnamefont
  {Sergienko}}\ and\ \bibinfo {author} {\bibfnamefont {E.}~\bibnamefont
  {Dagotto}},\ }\href {\doibase 10.1103/PhysRevB.73.094434} {\bibfield
  {journal} {\bibinfo  {journal} {Phys. Rev. B}\ }\textbf {\bibinfo {volume}
  {73}},\ \bibinfo {pages} {094434} (\bibinfo {year} {2006})}\BibitemShut
  {NoStop}%
\bibitem [{\citenamefont {Zhao}\ \emph {et~al.}(2003)\citenamefont {Zhao},
  \citenamefont {Wang}, \citenamefont {Xiang}, \citenamefont {Su},\ and\
  \citenamefont {Yu}}]{Zhao2003}%
  \BibitemOpen
  \bibfield  {author} {\bibinfo {author} {\bibfnamefont {J.~Z.}\ \bibnamefont
  {Zhao}}, \bibinfo {author} {\bibfnamefont {X.~Q.}\ \bibnamefont {Wang}},
  \bibinfo {author} {\bibfnamefont {T.}~\bibnamefont {Xiang}}, \bibinfo
  {author} {\bibfnamefont {Z.~B.}\ \bibnamefont {Su}}, \ and\ \bibinfo {author}
  {\bibfnamefont {L.}~\bibnamefont {Yu}},\ }\href {\doibase
  10.1103/PhysRevLett.90.207204} {\bibfield  {journal} {\bibinfo  {journal}
  {Phys. Rev. Lett.}\ }\textbf {\bibinfo {volume} {90}},\ \bibinfo {pages}
  {207204} (\bibinfo {year} {2003})}\BibitemShut {NoStop}%
\bibitem [{\citenamefont {Dmitrienko}\ \emph {et~al.}(2014)\citenamefont
  {Dmitrienko}, \citenamefont {Ovchinnikova}, \citenamefont {Collins},
  \citenamefont {Nisbet}, \citenamefont {Beutier}, \citenamefont {Kvashnin},
  \citenamefont {Mazurenko}, \citenamefont {Lichtenstein},\ and\ \citenamefont
  {Katsnelson}}]{Dmitrienko2014}%
  \BibitemOpen
  \bibfield  {author} {\bibinfo {author} {\bibfnamefont {V.~E.}\ \bibnamefont
  {Dmitrienko}}, \bibinfo {author} {\bibfnamefont {E.~N.}\ \bibnamefont
  {Ovchinnikova}}, \bibinfo {author} {\bibfnamefont {S.~P.}\ \bibnamefont
  {Collins}}, \bibinfo {author} {\bibfnamefont {G.}~\bibnamefont {Nisbet}},
  \bibinfo {author} {\bibfnamefont {G.}~\bibnamefont {Beutier}}, \bibinfo
  {author} {\bibfnamefont {Y.~O.}\ \bibnamefont {Kvashnin}}, \bibinfo {author}
  {\bibfnamefont {V.~V.}\ \bibnamefont {Mazurenko}}, \bibinfo {author}
  {\bibfnamefont {A.~I.}\ \bibnamefont {Lichtenstein}}, \ and\ \bibinfo
  {author} {\bibfnamefont {M.~I.}\ \bibnamefont {Katsnelson}},\ }\href
  {http://dx.doi.org/10.1038/nphys2859} {\bibfield  {journal} {\bibinfo
  {journal} {Nat Phys}\ }\textbf {\bibinfo {volume} {10}},\ \bibinfo {pages}
  {202} (\bibinfo {year} {2014})}\BibitemShut {NoStop}%
\bibitem [{\citenamefont {Nembach}\ \emph {et~al.}(2015)\citenamefont
  {Nembach}, \citenamefont {Shaw}, \citenamefont {Weiler}, \citenamefont
  {Jue},\ and\ \citenamefont {Silva}}]{Nembach2015}%
  \BibitemOpen
  \bibfield  {author} {\bibinfo {author} {\bibfnamefont {H.~T.}\ \bibnamefont
  {Nembach}}, \bibinfo {author} {\bibfnamefont {J.~M.}\ \bibnamefont {Shaw}},
  \bibinfo {author} {\bibfnamefont {M.}~\bibnamefont {Weiler}}, \bibinfo
  {author} {\bibfnamefont {E.}~\bibnamefont {Jue}}, \ and\ \bibinfo {author}
  {\bibfnamefont {T.~J.}\ \bibnamefont {Silva}},\ }\href
  {http://dx.doi.org/10.1038/nphys3418} {\bibfield  {journal} {\bibinfo
  {journal} {Nat Phys}\ }\textbf {\bibinfo {volume} {11}},\ \bibinfo {pages}
  {825} (\bibinfo {year} {2015})}\BibitemShut {NoStop}%
\bibitem [{\citenamefont {Gangadharaiah}\ \emph {et~al.}(2008)\citenamefont
  {Gangadharaiah}, \citenamefont {Sun},\ and\ \citenamefont
  {Starykh}}]{Gangadharaiah2008}%
  \BibitemOpen
  \bibfield  {author} {\bibinfo {author} {\bibfnamefont {S.}~\bibnamefont
  {Gangadharaiah}}, \bibinfo {author} {\bibfnamefont {J.}~\bibnamefont {Sun}},
  \ and\ \bibinfo {author} {\bibfnamefont {O.~A.}\ \bibnamefont {Starykh}},\
  }\href {\doibase 10.1103/PhysRevB.78.054436} {\bibfield  {journal} {\bibinfo
  {journal} {Phys. Rev. B}\ }\textbf {\bibinfo {volume} {78}},\ \bibinfo
  {pages} {054436} (\bibinfo {year} {2008})}\BibitemShut {NoStop}%
\bibitem [{\citenamefont {Povarov}\ \emph {et~al.}(2011)\citenamefont
  {Povarov}, \citenamefont {Smirnov}, \citenamefont {Starykh}, \citenamefont
  {Petrov},\ and\ \citenamefont {Shapiro}}]{Povarov2011}%
  \BibitemOpen
  \bibfield  {author} {\bibinfo {author} {\bibfnamefont {K.~Y.}\ \bibnamefont
  {Povarov}}, \bibinfo {author} {\bibfnamefont {A.~I.}\ \bibnamefont
  {Smirnov}}, \bibinfo {author} {\bibfnamefont {O.~A.}\ \bibnamefont
  {Starykh}}, \bibinfo {author} {\bibfnamefont {S.~V.}\ \bibnamefont {Petrov}},
  \ and\ \bibinfo {author} {\bibfnamefont {A.~Y.}\ \bibnamefont {Shapiro}},\
  }\href {\doibase 10.1103/PhysRevLett.107.037204} {\bibfield  {journal}
  {\bibinfo  {journal} {Phys. Rev. Lett.}\ }\textbf {\bibinfo {volume} {107}},\
  \bibinfo {pages} {037204} (\bibinfo {year} {2011})}\BibitemShut {NoStop}%
\bibitem [{\citenamefont {Fayzullin}\ \emph {et~al.}(2013)\citenamefont
  {Fayzullin}, \citenamefont {Eremina}, \citenamefont {Eremin}, \citenamefont
  {Dittl}, \citenamefont {van Well}, \citenamefont {Ritter}, \citenamefont
  {Assmus}, \citenamefont {Deisenhofer}, \citenamefont {von Nidda},\ and\
  \citenamefont {Loidl}}]{Fayzullin2013}%
  \BibitemOpen
  \bibfield  {author} {\bibinfo {author} {\bibfnamefont {M.~A.}\ \bibnamefont
  {Fayzullin}}, \bibinfo {author} {\bibfnamefont {R.~M.}\ \bibnamefont
  {Eremina}}, \bibinfo {author} {\bibfnamefont {M.~V.}\ \bibnamefont {Eremin}},
  \bibinfo {author} {\bibfnamefont {A.}~\bibnamefont {Dittl}}, \bibinfo
  {author} {\bibfnamefont {N.}~\bibnamefont {van Well}}, \bibinfo {author}
  {\bibfnamefont {F.}~\bibnamefont {Ritter}}, \bibinfo {author} {\bibfnamefont
  {W.}~\bibnamefont {Assmus}}, \bibinfo {author} {\bibfnamefont
  {J.}~\bibnamefont {Deisenhofer}}, \bibinfo {author} {\bibfnamefont
  {H.-A.~K.}\ \bibnamefont {von Nidda}}, \ and\ \bibinfo {author}
  {\bibfnamefont {A.}~\bibnamefont {Loidl}},\ }\href {\doibase
  10.1103/PhysRevB.88.174421} {\bibfield  {journal} {\bibinfo  {journal} {Phys.
  Rev. B}\ }\textbf {\bibinfo {volume} {88}},\ \bibinfo {pages} {174421}
  (\bibinfo {year} {2013})}\BibitemShut {NoStop}%
\bibitem [{\citenamefont {H\"alg}\ \emph {et~al.}(2014)\citenamefont {H\"alg},
  \citenamefont {Lorenz}, \citenamefont {Povarov}, \citenamefont {M\aa{}nsson},
  \citenamefont {Skourski},\ and\ \citenamefont {Zheludev}}]{Haelg2014}%
  \BibitemOpen
  \bibfield  {author} {\bibinfo {author} {\bibfnamefont {M.}~\bibnamefont
  {H\"alg}}, \bibinfo {author} {\bibfnamefont {W.~E.~A.}\ \bibnamefont
  {Lorenz}}, \bibinfo {author} {\bibfnamefont {K.~Y.}\ \bibnamefont {Povarov}},
  \bibinfo {author} {\bibfnamefont {M.}~\bibnamefont {M\aa{}nsson}}, \bibinfo
  {author} {\bibfnamefont {Y.}~\bibnamefont {Skourski}}, \ and\ \bibinfo
  {author} {\bibfnamefont {A.}~\bibnamefont {Zheludev}},\ }\href {\doibase
  10.1103/PhysRevB.90.174413} {\bibfield  {journal} {\bibinfo  {journal} {Phys.
  Rev. B}\ }\textbf {\bibinfo {volume} {90}},\ \bibinfo {pages} {174413}
  (\bibinfo {year} {2014})}\BibitemShut {NoStop}%
\bibitem [{\citenamefont {Mena}\ \emph {et~al.}(2014)\citenamefont {Mena},
  \citenamefont {Perry}, \citenamefont {Perring}, \citenamefont {Le},
  \citenamefont {Guerrero}, \citenamefont {Storni}, \citenamefont {Adroja},
  \citenamefont {R\"uegg},\ and\ \citenamefont {McMorrow}}]{Mena2014}%
  \BibitemOpen
  \bibfield  {author} {\bibinfo {author} {\bibfnamefont {M.}~\bibnamefont
  {Mena}}, \bibinfo {author} {\bibfnamefont {R.~S.}\ \bibnamefont {Perry}},
  \bibinfo {author} {\bibfnamefont {T.~G.}\ \bibnamefont {Perring}}, \bibinfo
  {author} {\bibfnamefont {M.~D.}\ \bibnamefont {Le}}, \bibinfo {author}
  {\bibfnamefont {S.}~\bibnamefont {Guerrero}}, \bibinfo {author}
  {\bibfnamefont {M.}~\bibnamefont {Storni}}, \bibinfo {author} {\bibfnamefont
  {D.~T.}\ \bibnamefont {Adroja}}, \bibinfo {author} {\bibfnamefont
  {C.}~\bibnamefont {R\"uegg}}, \ and\ \bibinfo {author} {\bibfnamefont
  {D.~F.}\ \bibnamefont {McMorrow}},\ }\href {\doibase
  10.1103/PhysRevLett.113.047202} {\bibfield  {journal} {\bibinfo  {journal}
  {Phys. Rev. Lett.}\ }\textbf {\bibinfo {volume} {113}},\ \bibinfo {pages}
  {047202} (\bibinfo {year} {2014})}\BibitemShut {NoStop}%
\bibitem [{\citenamefont {Blanchard}\ \emph {et~al.}(2015)\citenamefont
  {Blanchard}, \citenamefont {Sjolander}, \citenamefont {King}, \citenamefont
  {Ledbetter}, \citenamefont {Levine}, \citenamefont {Bajaj}, \citenamefont
  {Budker},\ and\ \citenamefont {Pines}}]{Blanchard2015}%
  \BibitemOpen
  \bibfield  {author} {\bibinfo {author} {\bibfnamefont {J.~W.}\ \bibnamefont
  {Blanchard}}, \bibinfo {author} {\bibfnamefont {T.~F.}\ \bibnamefont
  {Sjolander}}, \bibinfo {author} {\bibfnamefont {J.~P.}\ \bibnamefont {King}},
  \bibinfo {author} {\bibfnamefont {M.~P.}\ \bibnamefont {Ledbetter}}, \bibinfo
  {author} {\bibfnamefont {E.~H.}\ \bibnamefont {Levine}}, \bibinfo {author}
  {\bibfnamefont {V.~S.}\ \bibnamefont {Bajaj}}, \bibinfo {author}
  {\bibfnamefont {D.}~\bibnamefont {Budker}}, \ and\ \bibinfo {author}
  {\bibfnamefont {A.}~\bibnamefont {Pines}},\ }\href {\doibase
  10.1103/PhysRevB.92.220202} {\bibfield  {journal} {\bibinfo  {journal} {Phys.
  Rev. B}\ }\textbf {\bibinfo {volume} {92}},\ \bibinfo {pages} {220202}
  (\bibinfo {year} {2015})}\BibitemShut {NoStop}%
\bibitem [{\citenamefont {Hastings-Simon}\ \emph
  {et~al.}(2008{\natexlab{a}})\citenamefont {Hastings-Simon}, \citenamefont
  {Afzelius}, \citenamefont {J.Min\"{a}r}, \citenamefont {Staudt},
  \citenamefont {Lauritzen}, \citenamefont {de~Riedmatten}, \citenamefont
  {N.Gisin}, \citenamefont {A.Amari}, \citenamefont {Walther}, \citenamefont
  {Kröll}, \citenamefont {Cavalli},\ and\ \citenamefont
  {Bettinelli}}]{Hastings-Simon2008}%
  \BibitemOpen
  \bibfield  {author} {\bibinfo {author} {\bibfnamefont {S.~R.}\ \bibnamefont
  {Hastings-Simon}}, \bibinfo {author} {\bibfnamefont {M.}~\bibnamefont
  {Afzelius}}, \bibinfo {author} {\bibnamefont {J.Min\"{a}r}}, \bibinfo
  {author} {\bibfnamefont {M.}~\bibnamefont {Staudt}}, \bibinfo {author}
  {\bibfnamefont {B.}~\bibnamefont {Lauritzen}}, \bibinfo {author}
  {\bibfnamefont {H.}~\bibnamefont {de~Riedmatten}}, \bibinfo {author}
  {\bibnamefont {N.Gisin}}, \bibinfo {author} {\bibnamefont {A.Amari}},
  \bibinfo {author} {\bibfnamefont {A.}~\bibnamefont {Walther}}, \bibinfo
  {author} {\bibfnamefont {S.}~\bibnamefont {Kröll}}, \bibinfo {author}
  {\bibfnamefont {E.}~\bibnamefont {Cavalli}}, \ and\ \bibinfo {author}
  {\bibfnamefont {M.}~\bibnamefont {Bettinelli}},\ }\href@noop {} {\bibfield
  {journal} {\bibinfo  {journal} {Phys. Rev. B}\ }\textbf {\bibinfo {volume}
  {77}},\ \bibinfo {pages} {125111} (\bibinfo {year}
  {2008}{\natexlab{a}})}\BibitemShut {NoStop}%
\bibitem [{\citenamefont {Guillot-No\"el}\ \emph {et~al.}(2000)\citenamefont
  {Guillot-No\"el}, \citenamefont {Mehta}, \citenamefont {Viana}, \citenamefont
  {Gourier}, \citenamefont {Boukhris},\ and\ \citenamefont
  {Jandl}}]{Guillot-Noel2000}%
  \BibitemOpen
  \bibfield  {author} {\bibinfo {author} {\bibfnamefont {O.}~\bibnamefont
  {Guillot-No\"el}}, \bibinfo {author} {\bibfnamefont {V.}~\bibnamefont
  {Mehta}}, \bibinfo {author} {\bibfnamefont {B.}~\bibnamefont {Viana}},
  \bibinfo {author} {\bibfnamefont {D.}~\bibnamefont {Gourier}}, \bibinfo
  {author} {\bibfnamefont {M.}~\bibnamefont {Boukhris}}, \ and\ \bibinfo
  {author} {\bibfnamefont {S.}~\bibnamefont {Jandl}},\ }\href {\doibase
  10.1103/PhysRevB.61.15338} {\bibfield  {journal} {\bibinfo  {journal} {Phys.
  Rev. B}\ }\textbf {\bibinfo {volume} {61}},\ \bibinfo {pages} {15338}
  (\bibinfo {year} {2000})}\BibitemShut {NoStop}%
\bibitem [{\citenamefont {Macfarlane}\ and\ \citenamefont
  {Shelby}(1987)}]{Macfarlane1987a}%
  \BibitemOpen
  \bibfield  {author} {\bibinfo {author} {\bibfnamefont {R.}~\bibnamefont
  {Macfarlane}}\ and\ \bibinfo {author} {\bibfnamefont {R.}~\bibnamefont
  {Shelby}},\ }\href@noop {} {\emph {\bibinfo {title} {Coherent Transients And
  Holeburning Spectroscopy In Rare Earth Ions In Solids; Spectroscopy Of
  Crystals Containing Rare Earth Ions}}},\ edited by\ \bibinfo {editor}
  {\bibfnamefont {A.}~\bibnamefont {Kaplyankii}}\ and\ \bibinfo {editor}
  {\bibfnamefont {R.}~\bibnamefont {Macfarlane}}\ (\bibinfo  {publisher}
  {Elsevier Science Publishers},\ \bibinfo {address} {Amsterdam, Netherlands},\
  \bibinfo {year} {1987})\BibitemShut {NoStop}%
\bibitem [{\citenamefont {Hastings-Simon}\ \emph
  {et~al.}(2008{\natexlab{b}})\citenamefont {Hastings-Simon}, \citenamefont
  {Lauritzen}, \citenamefont {Staudt}, \citenamefont {van Mechelen},
  \citenamefont {Simon}, \citenamefont {de~Riedmatten}, \citenamefont
  {Afzelius},\ and\ \citenamefont {Gisin}}]{Hastings-Simon2008a}%
  \BibitemOpen
  \bibfield  {author} {\bibinfo {author} {\bibfnamefont {S.~R.}\ \bibnamefont
  {Hastings-Simon}}, \bibinfo {author} {\bibfnamefont {B.}~\bibnamefont
  {Lauritzen}}, \bibinfo {author} {\bibfnamefont {M.~U.}\ \bibnamefont
  {Staudt}}, \bibinfo {author} {\bibfnamefont {J.~L.~M.}\ \bibnamefont {van
  Mechelen}}, \bibinfo {author} {\bibfnamefont {C.}~\bibnamefont {Simon}},
  \bibinfo {author} {\bibfnamefont {H.}~\bibnamefont {de~Riedmatten}}, \bibinfo
  {author} {\bibfnamefont {M.}~\bibnamefont {Afzelius}}, \ and\ \bibinfo
  {author} {\bibfnamefont {N.}~\bibnamefont {Gisin}},\ }\href
  {http://link.aps.org/abstract/PRB/v78/e085410} {\bibfield  {journal}
  {\bibinfo  {journal} {Phys. Rev. B}\ }\textbf {\bibinfo {volume} {78}},\
  \bibinfo {pages} {085410} (\bibinfo {year} {2008}{\natexlab{b}})}\BibitemShut
  {NoStop}%
\bibitem [{\citenamefont {Afzelius}\ \emph {et~al.}(2010)\citenamefont
  {Afzelius}, \citenamefont {Staudt}, \citenamefont {de~Riedmatten},
  \citenamefont {Gisin}, \citenamefont {Guillot-No\"{e}l}, \citenamefont
  {Goldner}, \citenamefont {Marino}, \citenamefont {Porcher}, \citenamefont
  {Cavalli},\ and\ \citenamefont {Bettinelli}}]{Afzelius2010b}%
  \BibitemOpen
  \bibfield  {author} {\bibinfo {author} {\bibfnamefont {M.}~\bibnamefont
  {Afzelius}}, \bibinfo {author} {\bibfnamefont {M.~U.}\ \bibnamefont
  {Staudt}}, \bibinfo {author} {\bibfnamefont {H.}~\bibnamefont
  {de~Riedmatten}}, \bibinfo {author} {\bibfnamefont {N.}~\bibnamefont
  {Gisin}}, \bibinfo {author} {\bibfnamefont {O.}~\bibnamefont
  {Guillot-No\"{e}l}}, \bibinfo {author} {\bibfnamefont {P.}~\bibnamefont
  {Goldner}}, \bibinfo {author} {\bibfnamefont {R.}~\bibnamefont {Marino}},
  \bibinfo {author} {\bibfnamefont {P.}~\bibnamefont {Porcher}}, \bibinfo
  {author} {\bibfnamefont {E.}~\bibnamefont {Cavalli}}, \ and\ \bibinfo
  {author} {\bibfnamefont {M.}~\bibnamefont {Bettinelli}},\ }\href
  {http://www.sciencedirect.com/science/article/pii/S0022231309006280}
  {\bibfield  {journal} {\bibinfo  {journal} {Journal of Luminescence}\
  }\textbf {\bibinfo {volume} {130}},\ \bibinfo {pages} {1566} (\bibinfo {year}
  {2010})}\BibitemShut {NoStop}%
\bibitem [{\citenamefont {Ahlefeldt}\ \emph
  {et~al.}(2013{\natexlab{a}})\citenamefont {Ahlefeldt}, \citenamefont
  {Hutchison}, \citenamefont {Manson},\ and\ \citenamefont
  {Sellars}}]{Ahlefeldt2013a}%
  \BibitemOpen
  \bibfield  {author} {\bibinfo {author} {\bibfnamefont {R.~L.}\ \bibnamefont
  {Ahlefeldt}}, \bibinfo {author} {\bibfnamefont {W.~D.}\ \bibnamefont
  {Hutchison}}, \bibinfo {author} {\bibfnamefont {N.~B.}\ \bibnamefont
  {Manson}}, \ and\ \bibinfo {author} {\bibfnamefont {M.~J.}\ \bibnamefont
  {Sellars}},\ }\href {\doibase 10.1103/PhysRevB.88.184424} {\bibfield
  {journal} {\bibinfo  {journal} {Phys. Rev. B}\ }\textbf {\bibinfo {volume}
  {88}},\ \bibinfo {pages} {184424} (\bibinfo {year}
  {2013}{\natexlab{a}})}\BibitemShut {NoStop}%
\bibitem [{\citenamefont {Ahlefeldt}\ \emph
  {et~al.}(2013{\natexlab{b}})\citenamefont {Ahlefeldt}, \citenamefont
  {McAuslan}, \citenamefont {Longdell}, \citenamefont {Manson},\ and\
  \citenamefont {Sellars}}]{Ahlefeldt2013}%
  \BibitemOpen
  \bibfield  {author} {\bibinfo {author} {\bibfnamefont {R.~L.}\ \bibnamefont
  {Ahlefeldt}}, \bibinfo {author} {\bibfnamefont {D.~L.}\ \bibnamefont
  {McAuslan}}, \bibinfo {author} {\bibfnamefont {J.~J.}\ \bibnamefont
  {Longdell}}, \bibinfo {author} {\bibfnamefont {N.~B.}\ \bibnamefont
  {Manson}}, \ and\ \bibinfo {author} {\bibfnamefont {M.~J.}\ \bibnamefont
  {Sellars}},\ }\href {\doibase 10.1103/PhysRevLett.111.240501} {\bibfield
  {journal} {\bibinfo  {journal} {Phys. Rev. Lett.}\ }\textbf {\bibinfo
  {volume} {111}},\ \bibinfo {pages} {240501} (\bibinfo {year}
  {2013}{\natexlab{b}})}\BibitemShut {NoStop}%
\bibitem [{\citenamefont {Kubo}\ \emph {et~al.}(2010)\citenamefont {Kubo},
  \citenamefont {Ong}, \citenamefont {Bertet}, \citenamefont {Vion},
  \citenamefont {Jacques}, \citenamefont {Zheng}, \citenamefont {Dr\'eau},
  \citenamefont {Roch}, \citenamefont {Auffeves}, \citenamefont {Jelezko},
  \citenamefont {Wrachtrup}, \citenamefont {Barthe}, \citenamefont {Bergonzo},\
  and\ \citenamefont {Esteve}}]{Kubo2010}%
  \BibitemOpen
  \bibfield  {author} {\bibinfo {author} {\bibfnamefont {Y.}~\bibnamefont
  {Kubo}}, \bibinfo {author} {\bibfnamefont {F.~R.}\ \bibnamefont {Ong}},
  \bibinfo {author} {\bibfnamefont {P.}~\bibnamefont {Bertet}}, \bibinfo
  {author} {\bibfnamefont {D.}~\bibnamefont {Vion}}, \bibinfo {author}
  {\bibfnamefont {V.}~\bibnamefont {Jacques}}, \bibinfo {author} {\bibfnamefont
  {D.}~\bibnamefont {Zheng}}, \bibinfo {author} {\bibfnamefont
  {A.}~\bibnamefont {Dr\'eau}}, \bibinfo {author} {\bibfnamefont {J.-F.}\
  \bibnamefont {Roch}}, \bibinfo {author} {\bibfnamefont {A.}~\bibnamefont
  {Auffeves}}, \bibinfo {author} {\bibfnamefont {F.}~\bibnamefont {Jelezko}},
  \bibinfo {author} {\bibfnamefont {J.}~\bibnamefont {Wrachtrup}}, \bibinfo
  {author} {\bibfnamefont {M.~F.}\ \bibnamefont {Barthe}}, \bibinfo {author}
  {\bibfnamefont {P.}~\bibnamefont {Bergonzo}}, \ and\ \bibinfo {author}
  {\bibfnamefont {D.}~\bibnamefont {Esteve}},\ }\href {\doibase
  10.1103/PhysRevLett.105.140502} {\bibfield  {journal} {\bibinfo  {journal}
  {Phys. Rev. Lett.}\ }\textbf {\bibinfo {volume} {105}},\ \bibinfo {pages}
  {140502} (\bibinfo {year} {2010})}\BibitemShut {NoStop}%
\bibitem [{\citenamefont {Schuster}\ \emph {et~al.}(2010)\citenamefont
  {Schuster}, \citenamefont {Sears}, \citenamefont {Ginossar}, \citenamefont
  {DiCarlo}, \citenamefont {Frunzio}, \citenamefont {Morton}, \citenamefont
  {Wu}, \citenamefont {Briggs}, \citenamefont {Buckley}, \citenamefont
  {Awschalom},\ and\ \citenamefont {Schoelkopf}}]{Schuster2010}%
  \BibitemOpen
  \bibfield  {author} {\bibinfo {author} {\bibfnamefont {D.~I.}\ \bibnamefont
  {Schuster}}, \bibinfo {author} {\bibfnamefont {A.~P.}\ \bibnamefont {Sears}},
  \bibinfo {author} {\bibfnamefont {E.}~\bibnamefont {Ginossar}}, \bibinfo
  {author} {\bibfnamefont {L.}~\bibnamefont {DiCarlo}}, \bibinfo {author}
  {\bibfnamefont {L.}~\bibnamefont {Frunzio}}, \bibinfo {author} {\bibfnamefont
  {J.~J.~L.}\ \bibnamefont {Morton}}, \bibinfo {author} {\bibfnamefont
  {H.}~\bibnamefont {Wu}}, \bibinfo {author} {\bibfnamefont {G.~A.~D.}\
  \bibnamefont {Briggs}}, \bibinfo {author} {\bibfnamefont {B.~B.}\
  \bibnamefont {Buckley}}, \bibinfo {author} {\bibfnamefont {D.~D.}\
  \bibnamefont {Awschalom}}, \ and\ \bibinfo {author} {\bibfnamefont {R.~J.}\
  \bibnamefont {Schoelkopf}},\ }\href {\doibase 10.1103/PhysRevLett.105.140501}
  {\bibfield  {journal} {\bibinfo  {journal} {Phys. Rev. Lett.}\ }\textbf
  {\bibinfo {volume} {105}},\ \bibinfo {pages} {140501} (\bibinfo {year}
  {2010})}\BibitemShut {NoStop}%
\bibitem [{\citenamefont {Bushev}\ \emph {et~al.}(2011)\citenamefont {Bushev},
  \citenamefont {Feofanov}, \citenamefont {Rotzinger}, \citenamefont
  {Protopopov}, \citenamefont {Cole}, \citenamefont {Wilson}, \citenamefont
  {Fischer}, \citenamefont {Lukashenko},\ and\ \citenamefont
  {Ustinov}}]{Bushev2011}%
  \BibitemOpen
  \bibfield  {author} {\bibinfo {author} {\bibfnamefont {P.}~\bibnamefont
  {Bushev}}, \bibinfo {author} {\bibfnamefont {A.~K.}\ \bibnamefont
  {Feofanov}}, \bibinfo {author} {\bibfnamefont {H.}~\bibnamefont {Rotzinger}},
  \bibinfo {author} {\bibfnamefont {I.}~\bibnamefont {Protopopov}}, \bibinfo
  {author} {\bibfnamefont {J.~H.}\ \bibnamefont {Cole}}, \bibinfo {author}
  {\bibfnamefont {C.~M.}\ \bibnamefont {Wilson}}, \bibinfo {author}
  {\bibfnamefont {G.}~\bibnamefont {Fischer}}, \bibinfo {author} {\bibfnamefont
  {A.}~\bibnamefont {Lukashenko}}, \ and\ \bibinfo {author} {\bibfnamefont
  {A.~V.}\ \bibnamefont {Ustinov}},\ }\href
  {http://link.aps.org/doi/10.1103/PhysRevB.84.060501} {\bibfield  {journal}
  {\bibinfo  {journal} {Phys. Rev. B}\ }\textbf {\bibinfo {volume} {84}},\
  \bibinfo {pages} {060501} (\bibinfo {year} {2011})}\BibitemShut {NoStop}%
\bibitem [{\citenamefont {Clauss}\ \emph {et~al.}(2013)\citenamefont {Clauss},
  \citenamefont {Bothner}, \citenamefont {Koelle}, \citenamefont {Kleiner},
  \citenamefont {Bogani}, \citenamefont {Scheffler},\ and\ \citenamefont
  {Dressel}}]{Clauss2013}%
  \BibitemOpen
  \bibfield  {author} {\bibinfo {author} {\bibfnamefont {C.}~\bibnamefont
  {Clauss}}, \bibinfo {author} {\bibfnamefont {D.}~\bibnamefont {Bothner}},
  \bibinfo {author} {\bibfnamefont {D.}~\bibnamefont {Koelle}}, \bibinfo
  {author} {\bibfnamefont {R.}~\bibnamefont {Kleiner}}, \bibinfo {author}
  {\bibfnamefont {L.}~\bibnamefont {Bogani}}, \bibinfo {author} {\bibfnamefont
  {M.}~\bibnamefont {Scheffler}}, \ and\ \bibinfo {author} {\bibfnamefont
  {M.}~\bibnamefont {Dressel}},\ }\href
  {http://scitation.aip.org/content/aip/journal/apl/102/16/10.1063/1.4802956}
  {\bibfield  {journal} {\bibinfo  {journal} {Applied Physics Letters}\
  }\textbf {\bibinfo {volume} {102}},\ \bibinfo {eid} {162601} (\bibinfo {year}
  {2013})}\BibitemShut {NoStop}%
\bibitem [{\citenamefont {Wiemann}\ \emph {et~al.}(2015)\citenamefont
  {Wiemann}, \citenamefont {Simmendinger}, \citenamefont {Clauss},
  \citenamefont {Bogani}, \citenamefont {Bothner}, \citenamefont {Koelle},
  \citenamefont {Kleiner}, \citenamefont {Dressel},\ and\ \citenamefont
  {Scheffler}}]{Wiemann2015}%
  \BibitemOpen
  \bibfield  {author} {\bibinfo {author} {\bibfnamefont {Y.}~\bibnamefont
  {Wiemann}}, \bibinfo {author} {\bibfnamefont {J.}~\bibnamefont
  {Simmendinger}}, \bibinfo {author} {\bibfnamefont {C.}~\bibnamefont
  {Clauss}}, \bibinfo {author} {\bibfnamefont {L.}~\bibnamefont {Bogani}},
  \bibinfo {author} {\bibfnamefont {D.}~\bibnamefont {Bothner}}, \bibinfo
  {author} {\bibfnamefont {D.}~\bibnamefont {Koelle}}, \bibinfo {author}
  {\bibfnamefont {R.}~\bibnamefont {Kleiner}}, \bibinfo {author} {\bibfnamefont
  {M.}~\bibnamefont {Dressel}}, \ and\ \bibinfo {author} {\bibfnamefont
  {M.}~\bibnamefont {Scheffler}},\ }\href
  {http://scitation.aip.org/content/aip/journal/apl/106/19/10.1063/1.4921231}
  {\bibfield  {journal} {\bibinfo  {journal} {Applied Physics Letters}\
  }\textbf {\bibinfo {volume} {106}},\ \bibinfo {eid} {193505} (\bibinfo {year}
  {2015})}\BibitemShut {NoStop}%
\bibitem [{\citenamefont {Guillot-No\"{e}l}\ \emph {et~al.}(1999)\citenamefont
  {Guillot-No\"{e}l}, \citenamefont {Simons},\ and\ \citenamefont
  {Gourier}}]{Guillot-Noel1999}%
  \BibitemOpen
  \bibfield  {author} {\bibinfo {author} {\bibfnamefont {O.}~\bibnamefont
  {Guillot-No\"{e}l}}, \bibinfo {author} {\bibfnamefont {D.}~\bibnamefont
  {Simons}}, \ and\ \bibinfo {author} {\bibfnamefont {D.}~\bibnamefont
  {Gourier}},\ }\href@noop {} {\bibfield  {journal} {\bibinfo  {journal} {J.
  Phys. Chem. Sol.}\ }\textbf {\bibinfo {volume} {60}},\ \bibinfo {pages} {555}
  (\bibinfo {year} {1999})}\BibitemShut {NoStop}%
\bibitem [{\citenamefont {Suzuki}\ \emph {et~al.}(1980)\citenamefont {Suzuki},
  \citenamefont {Higashino},\ and\ \citenamefont {Inoue}}]{Suzuki1980}%
  \BibitemOpen
  \bibfield  {author} {\bibinfo {author} {\bibfnamefont {H.}~\bibnamefont
  {Suzuki}}, \bibinfo {author} {\bibfnamefont {Y.}~\bibnamefont {Higashino}}, \
  and\ \bibinfo {author} {\bibfnamefont {T.}~\bibnamefont {Inoue}},\
  }\href@noop {} {\bibfield  {journal} {\bibinfo  {journal} {Journal of the
  Physical Society of Japan}\ }\textbf {\bibinfo {volume} {49}},\ \bibinfo
  {pages} {1187} (\bibinfo {year} {1980})}\BibitemShut {NoStop}%
\bibitem [{\citenamefont {Suzuki}\ \emph {et~al.}(1983)\citenamefont {Suzuki},
  \citenamefont {Masuda},\ and\ \citenamefont {Miyamoto}}]{Suzuki1983}%
  \BibitemOpen
  \bibfield  {author} {\bibinfo {author} {\bibfnamefont {H.}~\bibnamefont
  {Suzuki}}, \bibinfo {author} {\bibfnamefont {Y.}~\bibnamefont {Masuda}}, \
  and\ \bibinfo {author} {\bibfnamefont {M.}~\bibnamefont {Miyamoto}},\
  }\href@noop {} {\bibfield  {journal} {\bibinfo  {journal} {Journal of the
  Physical Society of Japan}\ }\textbf {\bibinfo {volume} {52}},\ \bibinfo
  {pages} {250} (\bibinfo {year} {1983})}\BibitemShut {NoStop}%
\bibitem [{\citenamefont {Farr}\ \emph {et~al.}(2015)\citenamefont {Farr},
  \citenamefont {Goryachev}, \citenamefont {le~Floch}, \citenamefont {Bushev},\
  and\ \citenamefont {Tobar}}]{Farr2015}%
  \BibitemOpen
  \bibfield  {author} {\bibinfo {author} {\bibfnamefont {W.~G.}\ \bibnamefont
  {Farr}}, \bibinfo {author} {\bibfnamefont {M.}~\bibnamefont {Goryachev}},
  \bibinfo {author} {\bibfnamefont {J.-M.}\ \bibnamefont {le~Floch}}, \bibinfo
  {author} {\bibfnamefont {P.}~\bibnamefont {Bushev}}, \ and\ \bibinfo {author}
  {\bibfnamefont {M.~E.}\ \bibnamefont {Tobar}},\ }\href
  {http://scitation.aip.org/content/aip/journal/apl/107/12/10.1063/1.4931432}
  {\bibfield  {journal} {\bibinfo  {journal} {Applied Physics Letters}\
  }\textbf {\bibinfo {volume} {107}},\ \bibinfo {eid} {122401} (\bibinfo {year}
  {2015})}\BibitemShut {NoStop}%
\bibitem [{\citenamefont {Bussi\`{e}res}\ \emph {et~al.}(2013)\citenamefont
  {Bussi\`{e}res}, \citenamefont {Sangouard}, \citenamefont {Afzelius},
  \citenamefont {de~Riedmatten}, \citenamefont {Simon},\ and\ \citenamefont
  {Tittel}}]{Bussieres2013}%
  \BibitemOpen
  \bibfield  {author} {\bibinfo {author} {\bibfnamefont {F.}~\bibnamefont
  {Bussi\`{e}res}}, \bibinfo {author} {\bibfnamefont {N.}~\bibnamefont
  {Sangouard}}, \bibinfo {author} {\bibfnamefont {M.}~\bibnamefont {Afzelius}},
  \bibinfo {author} {\bibfnamefont {H.}~\bibnamefont {de~Riedmatten}}, \bibinfo
  {author} {\bibfnamefont {C.}~\bibnamefont {Simon}}, \ and\ \bibinfo {author}
  {\bibfnamefont {W.}~\bibnamefont {Tittel}},\ }\bibfield  {booktitle} \href {\doibase
  10.1080/09500340.2013.856482} {\bibfield  {journal} {\bibinfo  {journal}
  {Journal of Modern Optics}\ }\textbf {\bibinfo {volume} {60}},\ \bibinfo
  {pages} {1519} (\bibinfo {year} {2013})}\BibitemShut {NoStop}%
\bibitem [{\citenamefont {Afzelius}\ and\ \citenamefont
  {de~Riedmatten}(2015)}]{RiedmattenAfzeliusChapter2015}%
  \BibitemOpen
  \bibfield  {author} {\bibinfo {author} {\bibfnamefont {M.}~\bibnamefont
  {Afzelius}}\ and\ \bibinfo {author} {\bibfnamefont {H.}~\bibnamefont
  {de~Riedmatten}},\ }in\ \href {https://books.google.es/books?id=Fz5XrgEACAAJ}
  {\emph {\bibinfo {booktitle} {Engineering the Atom-Photon Interaction:
  Controlling Fundamental Processes with Photons, Atoms and Solids}}},\
  \bibinfo {series and number} {Nano-Optics and Nanophotonics},\ \bibinfo
  {editor} {edited by\ \bibinfo {editor} {\bibfnamefont {A.}~\bibnamefont
  {Predojevi\'{c}}}\ and\ \bibinfo {editor} {\bibfnamefont {M.~W.}\
  \bibnamefont {Mitchell}}}\ (\bibinfo  {publisher} {Springer International
  Publishing},\ \bibinfo {year} {2015})\ pp.\ \bibinfo {pages}
  {241--268}\BibitemShut {NoStop}%
\bibitem [{\citenamefont {Rippe}\ \emph {et~al.}(2008)\citenamefont {Rippe},
  \citenamefont {Julsgaard}, \citenamefont {Walther}, \citenamefont {Ying},\
  and\ \citenamefont {Kroll}}]{Rippe2008}%
  \BibitemOpen
  \bibfield  {author} {\bibinfo {author} {\bibfnamefont {L.}~\bibnamefont
  {Rippe}}, \bibinfo {author} {\bibfnamefont {B.}~\bibnamefont {Julsgaard}},
  \bibinfo {author} {\bibfnamefont {A.}~\bibnamefont {Walther}}, \bibinfo
  {author} {\bibfnamefont {Y.}~\bibnamefont {Ying}}, \ and\ \bibinfo {author}
  {\bibfnamefont {S.}~\bibnamefont {Kroll}},\ }\href
  {http://link.aps.org/abstract/PRA/v77/e022307} {\bibfield  {journal}
  {\bibinfo  {journal} {Phys. Rev. A}\ }\textbf {\bibinfo {volume} {77}},\
  \bibinfo {pages} {022307} (\bibinfo {year} {2008})}\BibitemShut {NoStop}%
\bibitem [{\citenamefont {Kolesov}\ \emph {et~al.}(2012)\citenamefont
  {Kolesov}, \citenamefont {Xia}, \citenamefont {Reuter}, \citenamefont
  {St\"{o}hr}, \citenamefont {Zappe}, \citenamefont {Meijer}, \citenamefont
  {Hemmer},\ and\ \citenamefont {Wrachtrup}}]{Kolesov2012}%
  \BibitemOpen
  \bibfield  {author} {\bibinfo {author} {\bibfnamefont {R.}~\bibnamefont
  {Kolesov}}, \bibinfo {author} {\bibfnamefont {K.}~\bibnamefont {Xia}},
  \bibinfo {author} {\bibfnamefont {R.}~\bibnamefont {Reuter}}, \bibinfo
  {author} {\bibfnamefont {R.}~\bibnamefont {St\"{o}hr}}, \bibinfo {author}
  {\bibfnamefont {A.}~\bibnamefont {Zappe}}, \bibinfo {author} {\bibfnamefont
  {J.}~\bibnamefont {Meijer}}, \bibinfo {author} {\bibfnamefont
  {P.}~\bibnamefont {Hemmer}}, \ and\ \bibinfo {author} {\bibfnamefont
  {J.}~\bibnamefont {Wrachtrup}},\ }\href
  {http://dx.doi.org/10.1038/ncomms2034} {\bibfield  {journal} {\bibinfo
  {journal} {Nat Commun}\ }\textbf {\bibinfo {volume} {3}},\ \bibinfo {pages}
  {1029} (\bibinfo {year} {2012})}\BibitemShut {NoStop}%
\bibitem [{\citenamefont {Yin}\ \emph {et~al.}(2013)\citenamefont {Yin},
  \citenamefont {Rancic}, \citenamefont {de~Boo}, \citenamefont {Stavrias},
  \citenamefont {McCallum}, \citenamefont {Sellars},\ and\ \citenamefont
  {Rogge}}]{Yin2013}%
  \BibitemOpen
  \bibfield  {author} {\bibinfo {author} {\bibfnamefont {C.}~\bibnamefont
  {Yin}}, \bibinfo {author} {\bibfnamefont {M.}~\bibnamefont {Rancic}},
  \bibinfo {author} {\bibfnamefont {G.~G.}\ \bibnamefont {de~Boo}}, \bibinfo
  {author} {\bibfnamefont {N.}~\bibnamefont {Stavrias}}, \bibinfo {author}
  {\bibfnamefont {J.~C.}\ \bibnamefont {McCallum}}, \bibinfo {author}
  {\bibfnamefont {M.~J.}\ \bibnamefont {Sellars}}, \ and\ \bibinfo {author}
  {\bibfnamefont {S.}~\bibnamefont {Rogge}},\ }\href
  {http://dx.doi.org/10.1038/nature12081} {\bibfield  {journal} {\bibinfo
  {journal} {Nature}\ }\textbf {\bibinfo {volume} {497}},\ \bibinfo {pages}
  {91} (\bibinfo {year} {2013})}\BibitemShut {NoStop}%
\bibitem [{\citenamefont {Siyushev}\ \emph {et~al.}(2014)\citenamefont
  {Siyushev}, \citenamefont {Xia}, \citenamefont {Reuter}, \citenamefont
  {Jamali}, \citenamefont {Zhao}, \citenamefont {Yang}, \citenamefont {Duan},
  \citenamefont {Kukharchyk}, \citenamefont {Wieck}, \citenamefont {Kolesov},\
  and\ \citenamefont {Wrachtrup}}]{Siyushev2014}%
  \BibitemOpen
  \bibfield  {author} {\bibinfo {author} {\bibfnamefont {P.}~\bibnamefont
  {Siyushev}}, \bibinfo {author} {\bibfnamefont {K.}~\bibnamefont {Xia}},
  \bibinfo {author} {\bibfnamefont {R.}~\bibnamefont {Reuter}}, \bibinfo
  {author} {\bibfnamefont {M.}~\bibnamefont {Jamali}}, \bibinfo {author}
  {\bibfnamefont {N.}~\bibnamefont {Zhao}}, \bibinfo {author} {\bibfnamefont
  {N.}~\bibnamefont {Yang}}, \bibinfo {author} {\bibfnamefont {C.}~\bibnamefont
  {Duan}}, \bibinfo {author} {\bibfnamefont {N.}~\bibnamefont {Kukharchyk}},
  \bibinfo {author} {\bibfnamefont {A.~D.}\ \bibnamefont {Wieck}}, \bibinfo
  {author} {\bibfnamefont {R.}~\bibnamefont {Kolesov}}, \ and\ \bibinfo
  {author} {\bibfnamefont {J.}~\bibnamefont {Wrachtrup}},\ }\href
  {http://dx.doi.org/10.1038/ncomms4895} {\bibfield  {journal} {\bibinfo
  {journal} {Nat Commun}\ }\textbf {\bibinfo {volume} {5}},\  (\bibinfo {year}
  {2014})}\BibitemShut {NoStop}%
\bibitem [{\citenamefont {Utikal}\ \emph {et~al.}(2014)\citenamefont {Utikal},
  \citenamefont {Eichhammer}, \citenamefont {Petersen}, \citenamefont {Renn},
  \citenamefont {G\"{o}tzinger},\ and\ \citenamefont
  {Sandoghdar}}]{Utikal2014}%
  \BibitemOpen
  \bibfield  {author} {\bibinfo {author} {\bibfnamefont {T.}~\bibnamefont
  {Utikal}}, \bibinfo {author} {\bibfnamefont {E.}~\bibnamefont {Eichhammer}},
  \bibinfo {author} {\bibfnamefont {L.}~\bibnamefont {Petersen}}, \bibinfo
  {author} {\bibfnamefont {A.}~\bibnamefont {Renn}}, \bibinfo {author}
  {\bibfnamefont {S.}~\bibnamefont {G\"{o}tzinger}}, \ and\ \bibinfo {author}
  {\bibfnamefont {V.}~\bibnamefont {Sandoghdar}},\ }\href
  {http://dx.doi.org/10.1038/ncomms4627} {\bibfield  {journal} {\bibinfo
  {journal} {Nat Commun}\ }\textbf {\bibinfo {volume} {5}},\  (\bibinfo {year}
  {2014})}\BibitemShut {NoStop}%
\bibitem [{\citenamefont {Xia}\ \emph {et~al.}(2015)\citenamefont {Xia},
  \citenamefont {Kolesov}, \citenamefont {Wang}, \citenamefont {Siyushev},
  \citenamefont {Reuter}, \citenamefont {Kornher}, \citenamefont {Kukharchyk},
  \citenamefont {Wieck}, \citenamefont {Villa}, \citenamefont {Yang},\ and\
  \citenamefont {Wrachtrup}}]{Xia2015}%
  \BibitemOpen
  \bibfield  {author} {\bibinfo {author} {\bibfnamefont {K.}~\bibnamefont
  {Xia}}, \bibinfo {author} {\bibfnamefont {R.}~\bibnamefont {Kolesov}},
  \bibinfo {author} {\bibfnamefont {Y.}~\bibnamefont {Wang}}, \bibinfo {author}
  {\bibfnamefont {P.}~\bibnamefont {Siyushev}}, \bibinfo {author}
  {\bibfnamefont {R.}~\bibnamefont {Reuter}}, \bibinfo {author} {\bibfnamefont
  {T.}~\bibnamefont {Kornher}}, \bibinfo {author} {\bibfnamefont
  {N.}~\bibnamefont {Kukharchyk}}, \bibinfo {author} {\bibfnamefont {A.~D.}\
  \bibnamefont {Wieck}}, \bibinfo {author} {\bibfnamefont {B.}~\bibnamefont
  {Villa}}, \bibinfo {author} {\bibfnamefont {S.}~\bibnamefont {Yang}}, \ and\
  \bibinfo {author} {\bibfnamefont {J.}~\bibnamefont {Wrachtrup}},\ }\href
  {\doibase 10.1103/PhysRevLett.115.093602} {\bibfield  {journal} {\bibinfo
  {journal} {Phys. Rev. Lett.}\ }\textbf {\bibinfo {volume} {115}},\ \bibinfo
  {pages} {093602} (\bibinfo {year} {2015})}\BibitemShut {NoStop}%
\bibitem [{\citenamefont {Eichhammer}\ \emph {et~al.}(2015)\citenamefont
  {Eichhammer}, \citenamefont {Utikal}, \citenamefont {G\"{o}tzinger},\ and\
  \citenamefont {Sandoghdar}}]{Eichhammer2015}%
  \BibitemOpen
  \bibfield  {author} {\bibinfo {author} {\bibfnamefont {E.}~\bibnamefont
  {Eichhammer}}, \bibinfo {author} {\bibfnamefont {T.}~\bibnamefont {Utikal}},
  \bibinfo {author} {\bibfnamefont {S.}~\bibnamefont {G\"{o}tzinger}}, \ and\
  \bibinfo {author} {\bibfnamefont {V.}~\bibnamefont {Sandoghdar}},\ }\href
  {http://stacks.iop.org/1367-2630/17/i=8/a=083018} {\bibfield  {journal}
  {\bibinfo  {journal} {New Journal of Physics}\ }\textbf {\bibinfo {volume}
  {17}},\ \bibinfo {pages} {083018} (\bibinfo {year} {2015})}\BibitemShut
  {NoStop}%
\bibitem [{\citenamefont {Petta}\ \emph {et~al.}(2005)\citenamefont {Petta},
  \citenamefont {Johnson}, \citenamefont {Taylor}, \citenamefont {Laird},
  \citenamefont {Yacoby}, \citenamefont {Lukin}, \citenamefont {Marcus},
  \citenamefont {Hanson},\ and\ \citenamefont {Gossard}}]{Petta2005}%
  \BibitemOpen
  \bibfield  {author} {\bibinfo {author} {\bibfnamefont {J.~R.}\ \bibnamefont
  {Petta}}, \bibinfo {author} {\bibfnamefont {A.~C.}\ \bibnamefont {Johnson}},
  \bibinfo {author} {\bibfnamefont {J.~M.}\ \bibnamefont {Taylor}}, \bibinfo
  {author} {\bibfnamefont {E.~A.}\ \bibnamefont {Laird}}, \bibinfo {author}
  {\bibfnamefont {A.}~\bibnamefont {Yacoby}}, \bibinfo {author} {\bibfnamefont
  {M.~D.}\ \bibnamefont {Lukin}}, \bibinfo {author} {\bibfnamefont {C.~M.}\
  \bibnamefont {Marcus}}, \bibinfo {author} {\bibfnamefont {M.~P.}\
  \bibnamefont {Hanson}}, \ and\ \bibinfo {author} {\bibfnamefont {A.~C.}\
  \bibnamefont {Gossard}},\ }\href
  {http://www.sciencemag.org/content/309/5744/2180.abstract} {\bibfield
  {journal} {\bibinfo  {journal} {Science}\ }\textbf {\bibinfo {volume}
  {309}},\ \bibinfo {pages} {2180} (\bibinfo {year} {2005})}\BibitemShut
  {NoStop}%
\bibitem [{\citenamefont {McAuslan}\ \emph {et~al.}(2012)\citenamefont
  {McAuslan}, \citenamefont {Bartholomew}, \citenamefont {Sellars},\ and\
  \citenamefont {Longdell}}]{McAuslan2012}%
  \BibitemOpen
  \bibfield  {author} {\bibinfo {author} {\bibfnamefont {D.~L.}\ \bibnamefont
  {McAuslan}}, \bibinfo {author} {\bibfnamefont {J.~G.}\ \bibnamefont
  {Bartholomew}}, \bibinfo {author} {\bibfnamefont {M.~J.}\ \bibnamefont
  {Sellars}}, \ and\ \bibinfo {author} {\bibfnamefont {J.~J.}\ \bibnamefont
  {Longdell}},\ }\href {\doibase 10.1103/PhysRevA.85.032339} {\bibfield
  {journal} {\bibinfo  {journal} {Phys. Rev. A}\ }\textbf {\bibinfo {volume}
  {85}},\ \bibinfo {pages} {032339} (\bibinfo {year} {2012})}\BibitemShut
  {NoStop}%
\end{thebibliography}
%

\bibliographystyle{apsrev4-1}

\end{document}